\documentstyle[preprint,aps]{revtex}
\tightenlines

\begin{document}
\title{Anomalous beating phase of the oscillating interlayer magnetoresistance in
layered metals}
\author{P.D. Grigoriev$^{1,2}$, M.V. Kartsovnik$^3$, W. Biberacher$^3$, N.D. Kushch$^4$ and P. Wyder$^{1}$}
\address{$^{1}$Grenoble High Magnetic Field Laboratory, MPI-FKF and CNRS, \\
BP 166, F-38042 Grenoble Cedex 09, France}
\address{$^{2}$L.D.Landau Institute for Theoretical Physics, \\
142432, Chernogolovka, Russia }
\address{$^3$ Walther-Mei\ss ner-Institut, Bayerische Akademie der Wissenschaften, \\
Walther-Mei\ss ner-Str. 8, D-85748, Germany}
\address{$^4$ Institute of Problems of Chemical Physics, 142432 Chernogolovka, Russia}
\date{\today }
\maketitle

\begin{abstract}
We analyze the beating behavior of the magnetic quantum oscillations in a
layered metal under the conditions when the cyclotron energy $\hbar \omega _c
$ is comparable to the interlayer transfer energy $t$. We find that the
positions of the beats in the interlayer resistance are considerably shifted
from those in the magnetization oscillations, and predict that the shift is
determined by the ratio $\hbar \omega _c/t$. A comparative study of the
Shubnikov-de Haas and de Haas-van Alphen effects in the
quasi-two-dimensional organic metal $\beta $-(BEDT-TTF)$_2$IBr$_2$ appears
to be consistent with the theoretical prediction.
\end{abstract}

\newpage In the last decade, the de Haas-van Alphen (dHvA) and Shubnikov-de
Haas (SdH) effects were extensively used for studying quasi-two-dimensional
(Q2D) organic metals (see for a review \cite{Wosnitzar,myrev,Single}). Due
to extremely high anisotropies of the electronic systems the amplitudes of
the oscillations are strongly enhanced in high-quality samples of these
materials (see e.g. \cite{Jerome,Laukhin}) and often cannot be described by
the Lifshitz-Kosevitch (LK) formula derived for conventional
three-dimensional metals \cite{Sh}. A number of theoretical works performed
in the last years on the dHvA effect in two-dimensional (2D) and Q2D systems 
\cite{Vagner,Harrison,IMV,GV,Champel,Pavel2} provide a consistent theory
which can be used for a quantitative analysis of experimental data obtained
on Q2D systems as long as many-body interactions only lead to constant
renormalization effects and magnetic breakdown effects are not concerned.

The situation with the Q2D SdH effect is more complicated; here even some
qualitative questions remain open. One of them is the origin of the phase
shift in the beats of the resistivity oscillations with respect to those in
the magnetization. The beating behavior of the oscillations in Q2D metals is
well known to originate from the slight warping of their Fermi surfaces in
the direction normal to the 2D plane. The superposition of the contributions
from the maximum and minimum cyclotron orbits is expected to lead to an
amplitude modulation of the $k$-th harmonic by the factor $\cos(2\pi k\Delta
F/2B-\pi /4)$, where $B$ is the magnetic field and $\Delta F =(c\hbar/2\pi
e) (A_{\text{max}}-A_{\text{min}})$ is the difference between the
oscillation frequencies caused by the extremal orbits with the $k$-space
areas $A_{\text{max}}$ and $A_{\text{min}}$, respectively \cite{Sh}. From
the beat frequency one can readily evaluate the warping of the Fermi surface
(FS) and hence the interlayer transfer integral, $4t \approx \epsilon_F
\Delta F/F$ (see e.g. \cite{Wosnitzar,SrRu}). The situation becomes less
clear when the warping is so weak that less than one half of the beat period
can be observed experimentally. In principle, an observation of one single
node would already be quite informative \cite{Weiss}, provided the phase
offset (i.e. the phase of the beat at $1/B\rightarrow 0$) is known. In the
standard LK theory this phase offset is strictly determined by geometrical
reasons to be equal to $-\pi /4$ for both dHvA and SdH effects \cite{Sh}.

However recent experiments on layered organic metals $\kappa $-(BEDT-TTF)$_2$
Cu[N(CN)$_2$]Br \cite{Weiss} and (BEDT-TTF)$_4$[Ni(dto)$_2$] \cite{Balthes}
have revealed a significant difference in the node positions of beating dHvA
and SdH signals. The respective phase shift in the latter compound was
estimated to be as big as $\pi /2$. Noteworthy, in both cases the
oscillation spectrum was strongly dominated by the first harmonic when no
substantial deviations from the standard LK theory is expected.

Although several potential reasons for this behavior have been outlined in
Refs.\cite{Weiss,Balthes} the most plausible one seems to be its association
with the Q2D nature of the electronic system \cite{Weiss}. However, the
limited amount of the reported experimental data is not sufficient for a
detailed analysis. Moreover, multiply connected FSs typical of both
compounds might lead to additional complications due to effects of magnetic
breakdown, interband scattering etc.

In order to clarify the problem, we have carried out comparative studies of
the oscillating magnetization and interlayer resistivity of the radical
cation salt $\beta $-(BEDT-TTF)$_2$IBr$_2$. This material exhibits a
relatively simple behavior without superstructure transitions or insulating
instabilities. It is normal metallic down to low temperatures and undergoes
a superconducting transition at $T_c \approx 2$ K \cite{IBr2}. Its
electronic properties are basically determined by a single cylindrical FS 
\cite{comment1} slightly (by $\simeq 1\%$) warped in the direction
perpendicular to the highly conducting BEDT-TTF layers \cite{MK1,Wosnitza}.
Thus, the present compound appears to be an ideal object for our purposes.
The experiment was done on a high-quality single crystal $\beta $-(BEDT-TTF)$
_2$IBr$_2$ at $T\approx 0.6$ K in magnetic field up to 16 T. To assure
exactly the same conditions for the dHvA and SdH effects (in particular,
identical field orientations are of crucial importance for our purposes!)
the measurements were performed in a set-up providing a simultaneous
registration of the magnetic torque and resistance \cite{Weiss}. In the
field range between 7 and 16 T we have observed clear beating with several
nodes in both dHvA and SdH signals. The positions of the beat nodes in the
SdH signal are found to be different from those in the dHvA signal, the
difference being dependent on the magnetic field.

Below we propose an explanation of the phenomenon based on the consideration
of both density of states (DoS) and Fermi velocity oscillations contributing
to the interlayer magnetotransport in a Q2D metal and then compare the
theoretical estimations with the experimental results.

We consider a Q2D metal in a magnetic field perpendicular to the conducting
layers with the energy spectrum 
\begin{equation}
\epsilon _{n,k_{z}}=\hbar \omega _{c}\,(n+1/2)-2t\cos (k_{z}d)  \label{ES}
\end{equation}
where $t$ is the interlayer transfer integral, $k_{z}$ is the wavevector
perpendicular to the layers, $d$ is the interlayer distance, $\omega
_{c}=eB/m^{\ast }c$ is the cyclotron frequency. Both $\hbar \omega _c$ and $t
$ are assumed to be much smaller than the Fermi energy.

The DoS of electron gas with this spectrum can be easily obtained performing
the summation over all quantum numbers at a fixed energy: 
\begin{equation}
g(\epsilon )=\sum_{n=0}^{\infty } \frac{N_{LL}}{\sqrt{4t^{2}-\left( \epsilon
-\hbar \omega _{c}\left( n+1/2\right) \right) ^{2}} }  \label{DoS0}
\end{equation}
where $N_{LL}$ is the Landau level degeneracy. The sum over Landau levels
can be represented as a harmonic series using the Poisson summation formula
\cite{ZW}.  As a result one gets\cite{Champel}: 
\begin{equation}
g(\epsilon ) \propto 1+2\sum_{k=1}^{\infty }(-1)^{k}\cos \left( \frac{2\pi
k\epsilon }{\hbar \omega _{c}}\right) J_{0}\left( \frac{4\pi kt}{\hbar
\omega _{c}}\right)  \label{DoS1}
\end{equation}

We shall consider the case $4 \pi t > \hbar \omega_{c}$ when the beats can
be observed. Then the zeroth order Bessel function $J_{0}\left( \pi k
4t/\hbar \omega _{c}\right)$ describing the beating of the DoS oscillations
can be simplified, as $J_{0}(x)\approx \sqrt{2/\pi x}\cos ( x-\pi /4)$.
Further, we consider the limit of strong harmonic damping, retaining only
the zeroth and first harmonics. In this limit the oscillations of the
chemical potential $\mu $ can be neglected. Knowing the DoS we can now
obtain the oscillating part of the magnetization as \cite{Champel} 
\begin{equation}
\tilde M \propto \sin \left( \frac{2\pi \mu }{\hbar \omega _{c}}\right) \cos
\left( \frac{4 \pi t}{\hbar \omega _{c}}-\frac{\pi }{4}\right) R_{T}
\label{tM}
\end{equation}
where $R_T$ is the usual temperature smearing factor. This expression
coincides with the result of the three-dimensional LK theory \cite{Sh} and
allows an evaluation of $t$ from the beat frequency.

The interlayer conductivity $\sigma _{zz}$ can be approximately evaluated
from the Boltzmann transport equation\cite{Mah}, assuming the impurities are
point-like: 
\begin{equation}
\sigma _{zz}=e^{2}\sum_{m\equiv \{n,k_{z},k_{x}\}} v_{z}^{2} (k_{z})\,\delta
(\epsilon (n,k_{z})-\mu ) \cdot \tau (\mu ) \equiv e^{2} I(\mu ) \tau (\mu )
\label{DF}
\end{equation}
where $v_{z}$ is the z-component of the electron velocity and $\tau (\mu )$
the momentum relaxation time at the Fermi level. The latter, in Born
approximation, is inversely proportional to the DoS: $1/\tau (\mu ) \propto
g(\mu )$ and oscillates in magnetic field according to Eq.(\ref{DoS1}). In
addition, when the cyclotron energy is comparable to the warping of the FS,
the oscillations of the electron velocity summed over the states at the
Fermi level, $I(\mu ) \equiv \sum_{\epsilon=\mu }\left| v_{z}\right| ^{2}$
become also important \cite{Harrison}.

To calculate this quantity one has to perform the integrations over $k_{x}$
and $k_{z}$ in (\ref{DF}) and substitute the expression for $v_{z}$: 
\begin{equation}
v_{z}(\epsilon ,n) =\frac{d}{\hbar }\sqrt{4t^{2}-\left( \epsilon -\hbar
\omega _{c}\left( \,n+1/2\right) \right) ^{2}}  \label{vz}
\end{equation}
As a result one obtains 
\[
I(\epsilon ) =\sum_{n=0}^{\infty } \frac{N_{LL}d^2}{2\pi \hbar } \sqrt{
4t^{2}-\left( \epsilon -\hbar \omega _{c}\left( n+1/2 \right) \right) ^{2}}
= \frac{4N_{LL}d^{2}t}{\pi \hbar ^{2}}\times 
\]
\begin{equation}
\times \left[ \frac{4 \pi t}{8\hbar \omega _{c}}+\sum_{k=1}^{\infty } 
\frac{ 
\left( -1\right) ^{k}}{2k}\cos \left( \frac{2\pi k\epsilon }{\hbar \omega
_{c}}\right) J_{1}\left( \frac{4\pi kt}{\hbar \omega _{c}}\right) \right]
\label{Ip}
\end{equation}
Here we again applied the Poisson summation formula \cite{ZW}.

The first order Bessel function $J_{1}( 4\pi kt/\hbar \omega _{c}) $
entering (\ref{Ip}) also describes beatings but with a phase different from
that of the DoS beatings given by $J_{0}(4 \pi kt/\hbar \omega _{c}) $ in
Eq.(\ref{DoS1}): at large $x$, $J_1(x)\approx \sqrt{2/\pi x} \sin (x-\pi /4)$
is just shifted by $\pi /2$ with respect to $J_0(x)$.

The phase shift in the beating of the oscillations in $g(\epsilon )$ and 
$I(\epsilon )$ is illustrated in Fig. 1 in which these quantities are plotted
for two different values of the ratio $4t/\hbar \omega _c$. When $4t/\hbar
\omega _c=2.25$ the DoS oscillations have a maximum amplitude of the first
harmonic (Fig. 1a). At the same time the oscillations of $I(\epsilon )$
exhibit a nearly zero amplitude of the first harmonic as shown on Fig. 1c
whilst their second harmonic (not shown in the figure) is at the maximum. By
contrast, when $4t/\hbar \omega _c=1.8$ (Fig. 1b,d) the first harmonic of
the DoS is at the node whereas that of $I(\epsilon )$ has the maximum
amplitude.

The difference between the phases of the beats in oscillating $g(\epsilon )$
and $I(\epsilon )$ leads to a shift of the beat phase of the SdH
oscillations with respect to that of the dHvA oscillations. Indeed, taking
into account that $\tau \propto 1/g$, substituting Eqs. (\ref{DoS1}) and 
(\ref{Ip}) into (\ref{DF}), applying the large argument expansions of $J_0(x)$
and $J_1(x)$ and introducing the temperature smearing, we come to the
following expression for the first harmonic of the interlayer conductivity: 
\begin{equation}
\tilde{\sigma}_{zz}\propto \cos \left( \frac{2\pi \mu }{\hbar \omega _{c}}
\right) \cos \left( \frac{4 \pi t}{\hbar \omega _{c}}-\frac{\pi }{4}+\phi
\right) R_{T}  \label{Sig7}
\end{equation}
where 
\begin{equation}
\phi \equiv \arctan (a) \text{\ and\ } a\equiv \hbar \omega _{c}/2 \pi t
\label{PS}
\end{equation}
Comparing these expressions with Eq.(\ref{tM}), one can see that the beats
in the SdH and dHvA oscillations can become considerably shifted with
respect to each other as the cyclotron frequency approaches the value of the
interlayer transfer integral.

The above evaluation is based on the semi-classical Boltzmann equation and
certainly is not expected to give a precise result. Nevertheless, as will be
seen below, its predictions concerning the phase shift are in good
qualitative agreement with the experiment and we therefore believe that it
correctly reflects the physics of the phenomenon.

Fig. 2 shows the oscillating parts of the magnetization and interlayer
magnetoresistance in the normal state of $\beta $-(BEDT-TTF)$_2$IBr$_2$ at
magnetic field tilted by $\theta \approx 14.8^{\circ }$ from the normal to
the BEDT-TTF layers. The curves have been obtained by subtracting slowly
varying backgrounds from the measured magnetic torque $\tau (B)$ and
resistance $R(B)$ and (for the magnetization oscillations, $\tilde M \propto
\tau /B$) subsequent dividing by $B$. The fast Fourier transformation (FFT)
spectra shown in the insets reveal the fundamental frequency of $\approx
3930 $ T in agreement with previous works \cite{MK1,Wosnitza}. The second
harmonic contribution is about $1\% $ of that from the fundamental one at
the highest field.

Clear beats with four nodes (indicated by arrows) are seen in both the dHvA
and SdH curves. We have assured that the observed beats originate from the
warping of the cylindrical FS by checking the angular dependence of their
frequency \cite{Wosnitza}. The fact that the oscillation amplitude does not
exactly vanish at the nodes was attributed to slightly different cyclotron
masses at the extremal orbits of the FS \cite{Wosnitza}. The positions of
the nodes determined as midpoints of narrow field intervals at which the
oscillations inverse the phase are plotted on Fig. 3. The straight line is a
linear fit of the magnetization data revealing the beat frequency $\Delta F
= 40.9$ T that, according to Eq.(\ref{tM}), corresponds to $4t/\epsilon _F =
\Delta F/F = 1/96$. The error bars in the node positions do not exceed $\pm
3\times 10^{-4}$ T$^{-1}$ for $N=3 \text{ to } 5$ and are somewhat bigger, 
$\simeq \pm 10^{-3}$ T$^{-1}$, for $N=6$ due to a lower signal-to-noise
ratio. We note that although the angle $\theta =14.8^{\circ }$ corresponds
to a region in the vicinity of the maximum beat frequency ($\max \{\Delta
F(\theta )\} \approx 42.0$ T for the given field rotation plane), the
sensitivity of the node positions to the field orientation is still quite
high: the nodes shift by $\approx 2.3 \times 10^{-3} \text{ T}^{-1}$ at
changing $\theta $ by $1^{\circ }$. Thus, if one has to remount the sample
between the torque and resistance measurements, even a slight misalignment
may cause a substantial additional error. In our experiment both quantities
were measured at the same field sweep, hence, such an error was eliminated.

From Figs. 2 and 3 one can see that the nodes of the SdH oscillations are
considerably shifted to higher fields with respect to those of the dHvA
oscillations. The shift grows with increasing the field. Both these
observations are fully consistent with the above theoretical prediction.

In order to make a further comparison between the experiment and theory, we
plot the quantity $\tan(\phi)$ (where $\phi$ is the phase shift between the
beating of the SdH and dHvA oscillations obtained from Fig. 3) as a function
of magnetic field in Fig. 4. A linear fit to this plot (dashed line in Fig.
4) has a slope of 0.037 1/T \cite{comment2}. A substitution of this value
and the cyclotron mass $m^{\ast }=4.2m_{\text e}$, obtained from the
temperature dependent amplitude of the fundamental harmonic, into Eq.(\ref
{PS}) yields an estimation for the interlayer bandwidth $4t \simeq 0.48$ meV
or the ratio $4t/\epsilon _F = \Delta F/F \simeq 1/230$. This is somewhat
smaller than the value 1/96 obtained directly from the ratio between the
beating and fundamental frequencies. However, taking into account an
approximate character of the presented theoretical model, the difference is
not surprising. Further theoretical work is needed in order to provide a
more explicit basis for the quantitative description of the phenomenon.

Summarizing, the beats of the SdH oscillations in $\beta $-(BEDT-TTF)$_2$IBr
$_2$ are found to be shifted towards higher fields with respect to those of
the dHvA signal. We attribute this effect to interfering contributions from
oscillating DoS and Fermi velocity to the interlayer conductivity of this
layered compound. Thus, the observed behavior appears to be a general
feature of Q2D metals which should be taken into account whenever the
cyclotron energy becomes comparable to the interlayer transfer energy.

We are thankful to A.M. Dyugaev, I. Vagner and A.E. Kovalev for stimulating
discussions. The work was supported by the EU ICN contract
HPRI-CT-1999-40013, and grants DFG-RFBR No. 436 RUS 113/592 and RFBR No.
00-02-17729a.

\newpage

\begin{center}
Figure captions
\end{center}

Fig. 1. a - the DoS (solid line) near the Fermi level and its first harmonic
(dashed line) according to Eq.(\ref{DoS1}), at the ratio $4t/\hbar \omega _c
= 2.25$; b - the same at the ratio $4t/\hbar \omega _c = 1.8$; c - the
quantity $I(\epsilon )$ (solid line) and its first harmonic (dashed line)
according to Eq.(7), at $4t/\hbar \omega _c = 2.25$; c - the same at 
$4t/\hbar \omega _c = 1.8$. In all four panels: the dotted lines are the
contributions from individual Landau levels.

Fig. 2. dHvA (left scale) and SdH (right scale) oscillations in $\beta 
$-(BEDT-TTF)$_2$IBr$_2$ at $\theta \approx 14.8^{\circ }$. Insets:
corresponding FFT spectra.

Fig. 3. The positions of the nodes in the oscillating magnetization (filled
symbols) and resistance (open symbols) versus inverse field. The straight
line is the linear fit to the magnetization data.

Fig. 4. Tangent of the phase shift between the node positions in the SdH and
dHvA signals taken from the data on Fig. 3 as a function of magnetic field.
The dashed line is a linear fit according to Eq.(\ref{PS}) \cite{comment2}.


\begin{references}
\bibitem{comment1}  An observation of slow oscillations of the resistivity
resembling strongly the SdH effect have lead to a suggestion of an
additional, very small FS pockets in this compound \cite{MK1}. However, our
recent analysis \cite{SO} have shown that such oscillations may originate
from the strongly anisotropic character of the main FS cylinder and do not
necessarily invoke additional pockets.
\bibitem{comment2}  For the fit shown on Fig. 4 only the data corresponding
to the nodes with $N=3\text{ to }5$ were used. The phase shift for $N=6$ was
disregarded due to the very large error bar.
\end{references}
\end{document}